\documentclass[twocolumn,showpacs,preprintnumbers,amsmath,amssymb]{revtex4}

\usepackage{graphicx}
\usepackage{dcolumn}
\usepackage{bm}
\usepackage{color}

\begin{document}

\preprint{}

\title{Enhancement of the CP-odd effect in the nuclear electric dipole moment of $^6$Li}

\author{Nodoka~Yamanaka$^1$}
  \email{nodoka.yamanaka@riken.jp}
\author{Emiko~Hiyama$^2$}
  \affiliation{$^1$iTHES Research Group, RIKEN, 
  Wako, Saitama 351-0198, Japan}
  \affiliation{$^2$RIKEN Nishina Center, RIKEN, 
  Wako, Saitama 351-0198, Japan}

\date{\today}

\begin{abstract}

We calculate for the first time the electric dipole moment (EDM) of the $^6$Li nucleus within the $\alpha + p + n$ three-body cluster model using the Gaussian expansion method, assuming the one meson exchange P, CP-odd nuclear forces.
It is found that the EDM of the $^6$Li is 2 times more sensitive on the isovector pion exchange P, CP-odd nuclear force than the deuteron EDM, due to the CP-odd interaction between the nucleons and the $\alpha$ cluster.
The $^9$Be EDM is also calculated in the same framework as an $\alpha + \alpha + n$ three-body system.
We also test the ab initio calculation of the EDM of the deuteron, $^3$H and $^3$He nuclei using the realistic Argonne $v18$ nuclear force.
In the ab initio calculations, good agreements with previous studies are obtained.
We finally discuss the prospects for the new physics beyond the standard model.

\end{abstract}


\maketitle

\section{Introduction}

It is commonly believed that the current baryon number asymmetry was created at the early stage of our Universe.
It is however known that the standard model of particle physics is not able to realize it, and CP violation beyond the standard model (BSM) is therefore required.
One promising experimental approach to search for the CP violation is the measurement of the {\it electric dipole moment} (EDM) \cite{hereview,Bernreuther,khriplovichbook,ginges,pospelovreview,fukuyama,engel,yamanaka,hewett}.
The EDM is an observable very sensitive to it, and has many advantages, such as the accurate measurability in experiments, low cost in the experimental preparation, very small standard model contribution \cite{smedm}, etc, and it was so far measured in various systems, such as the neutron \cite{baker}, atoms \cite{griffith}, molecules \cite{acme}, and muon \cite{muong2}.

Among them, the nuclear EDM is of particular interest \cite{hewett,storage}.
Recently, the measurement of the EDM of light nuclei using storage rings is planned at the Brookhaven National Laboratory, and the prospective sensitivity is $O(10^{-29}) e$ cm \cite{bnl}.
The nuclear EDM also has its own advantages.
First, the nuclear EDM does not suffer from Schiff's screening \cite{schiff}, since we have no electrons to screen the charged nucleus.
Also, the nuclear system may enhance the nucleon level CP violation due to the many-body effect \cite{sushkov}.
These arguments make the nuclear EDM to be a very attractive probe of strong sector CP violation, and theoretical investigations of the deuteron \cite{korkin,liu,pospelovdeuteron,afnan,devries,dedmtheta} and three-nucleon systems \cite{stetcu,chiral3nucleon,song,bsaisou,bsaisou2} were extensively done so far.

As a natural choice of the next target of the discussion, we have the $^6$Li nucleus, since it is the lightest stable system with non-zero angular momentum which can be found after the three-nucleon systems.
The $^6$Li nucleus is known to have a cluster structure \cite{hasegawa}, so we may expect some enhancement of the nucleon level CP violation, due to the derivative interaction of the CP-odd Hamiltonian.
In this sense, the analysis of the $^6$Li EDM may also be interesting in the point-of-view of the study of the nuclear structure and the cluster dynamics \cite{clusterreview}.

In this work, we therefore investigate the EDM of the $^6$Li nucleus in the cluster approximation, as the $\alpha - p - n$ three-body system.
We also calculate the EDM of the $^9$Be nucleus, which is calculable in the same framework \cite{arai}, and a test of ab initio calculations of the deuteron, the $^3$He, and $^3$H nuclei using the Argonne $v18$ interaction \cite{av18}.
To solve the many-body Schr\"{o}dinger equation, we use the Gaussian expansion method \cite{hiyama}.
This method was applied in a wide number of subjects, extending from the particle to atomic physics \cite{3nucleon,benchmark,gemapplication,Hiyama2012ptep}, and it is also expected to give accurate results in the study of the EDM of few-nucleon systems.

This paper is organized as follows.
We first give the definition of the EDM.
We then briefly review our calculational method.
The result of the EDM of the $^6$Li nucleus is then given, together with the EDMs of $^9$Be, $^3$He, $^3$H, and $^2$H nuclei.
We finally see the prospect for the determination of the new physics BSM and summarize our discussion.

\section{The nuclear electric dipole moment}

To induce the nuclear EDM, the existence of the P, CP-odd nucleon level processes is required.
The nuclear EDM has two leading sources: (1) the intrinsic EDM of the constituent nucleons, and (2) the P, CP-odd $N-N$ interactions (nuclear force) to polarize the whole nucleus.
In this work, we neglect the exchange current, since its contribution is expected to be small \cite{liu}.

We first give the contribution of the intrinsic EDM of the constituent nucleons to the nuclear EDM.
As the nucleon EDM is proportional to the nucleon spin, the effect is simply given by
\begin{eqnarray}
d_A^{\rm (Nedm)} 
&=&
\sum_{i}^A
d_i \langle \,A\, |\, \sigma_{iz} \, |\, A\, \rangle
\nonumber\\
&\equiv &
\langle \sigma_p \rangle_A \, d_p + \langle \sigma_n \rangle_A \, d_n
,
\end{eqnarray}
where $|\, A\, \rangle$ is the polarized nuclear wave function ($A=$ $^2$H, $^3$He, $^3$H, $^6$Li, $^9$Be). Here $d_p$ and $d_n$ are the proton and neutron EDMs, respectively.
These are given parameters which depend on the QCD and elementary level physics.
The coefficients $\langle \sigma_p \rangle_A$ and $\langle \sigma_n \rangle_A$ are just the spin matrix elements of the nucleus, and depend only on the nuclear structure.
The contribution of the single nucleon EDM may be enhanced if the nucleon is relativistic inside the nucleus, as for the atomic systems \cite{khriplovichbook,ginges,sandars}.
As the nucleons are nonrelativistic in light nuclei, the linear coefficients of the systems in question will not receive sizable enhancement.

The effect of the nuclear polarization generated by the CP-odd nuclear force may, in contrast, be enhanced even in light nuclear systems.
The polarization contribution of the P, CP-odd nuclear force to the nuclear EDM is given by
\begin{eqnarray}
d_{A}^{\rm (pol)} 
&=&
\sum_{i=1}^{A} \frac{e}{2} 
\langle \, \tilde A \, |\, (1+\tau_i^z ) \, {\cal R}_{iz} \, | \, \tilde A \, \rangle
,
\end{eqnarray}
where $|\, \tilde A\, \rangle$ is the polarized (in the $z$-axis) nuclear wave function, and $\tau^z_i$ is the isospin Pauli matrix.
${\cal R}_{iz}$ is (the $z$-component of) the position of the constituent nucleon in the nuclear center of mass frame.
This permanent polarization effect is realized through the mixing of opposite parity states.


We now show the detail of the CP-odd nuclear force, necessary to generate the polarization.
In this study, the CP-odd nuclear force is given by the standard one, based on one-meson exchange potential \cite{pvcpvhamiltonian}.
The CP-odd hamiltonian is given by
\begin{eqnarray}
H_{P\hspace{-.5em}/\, T\hspace{-.5em}/\, } 
& = &
 \frac{1}{2m_N} \bigg\{\bm\sigma_{-}\cdot\bm\nabla(\bar{G}_{\eta}^{(0)}\,\mathcal{Y}_{\eta}(r)-\bar{G}_{\omega}^{(0)}\,\mathcal{Y}_{\omega}(r))
 \nonumber \\
&& \hspace{3em}
+\bm\tau_{1}\cdot\bm\tau_{2}\,\bm\sigma_{-}\cdot\bm\nabla
\bigl[\bar{G}_{\pi}^{(0)}\,\mathcal{Y}_{\pi}(r)-\bar{G}_{\rho}^{(0)}\,\mathcal{Y}_{\rho}(r) \bigr]
\nonumber \\
&& \hspace{3em} 
+\frac{1}{2} \tau_{+}^{z}\,\bm\sigma_{-}\cdot\bm\nabla
\bigl[ \bar{G}_{\pi}^{(1)}\,\mathcal{Y}_{\pi}(r)-\bar{G}_{\eta}^{(1)}\,\mathcal{Y}_{\eta}(r)
\nonumber \\
&& \hspace{9.5em} 
-\bar{G}_{\rho}^{(1)}\,\mathcal{Y}_{\rho}(r)-\bar{G}_{\omega}^{(1)}\,\mathcal{Y}_{\omega}(r)
\bigr]
\nonumber \\
&& \hspace{3em}
+\frac{1}{2}\tau_{-}^{z}\,\bm\sigma_{+}\cdot\bm\nabla
\bigl[ \bar{G}_{\pi}^{(1)}\,\mathcal{Y}_{\pi}(r)+\bar{G}_{\eta}^{(1)}\,\mathcal{Y}_{\eta}(r)
\nonumber \\
&& \hspace{9.5em} 
+\bar{G}_{\rho}^{(1)}\,\mathcal{Y}_{\rho}(r)-\bar{G}_{\omega}^{(1)}\,\mathcal{Y}_{\omega}(r) \bigr]
\nonumber \\
&& \hspace{3em}
+(3\tau_{1}^{z}\tau_{2}^{z}-\bm\tau_{1}\cdot\bm\tau_{2})\,\bm\sigma_{-}\cdot\bm\nabla
\bigl[ \bar{G}_{\pi}^{(2)}\,\mathcal{Y}_{\pi}(r)
\nonumber \\
&& \hspace{13em} 
-\bar{G}_{\rho}^{(2)}\,\mathcal{Y}_{\rho}(r) \bigr]
\bigg\}
,
\label{eq:CPVhamiltonian}
\end{eqnarray}
where $\bar{G}_{X}^{(i)} \equiv g_{X NN} \bar g_{X NN}^{(i)}$ is the coupling constant of the CP-odd nuclear force with the exchanged mesons $X=\pi , \eta , \rho , \omega$.
The index $i=0,1,2$ denotes the isoscalar, isovector, and isotensor structures.
The Yukawa function is given by ${\cal Y}_X (r) =  \frac{e^{-m_X r }}{4 \pi r}$.
As the CP-odd effect is small, the polarization contribution to the nuclear EDM is given by the linear term
\begin{equation}
d_A^{\rm (pol)} 
=
\sum_{X,i} a_{A,X}^{(i)} \bar G_X^{(i)} 
.
\end{equation}
The linear coefficients $a_{A,X}^{(i)}$ depend only on the nuclear structure, and are the main targets of this discussion.


\section{Methodology and model setup}

We study the structure of the $^6$Li and $^9$Be nuclei within the framework of the $\alpha +n+p$ and 
$\alpha +\alpha +n$ three-body cluster models.
Here we assume the $\alpha$ cluster to be an inert core.
Also, we take three kinds of Jacobian coordinates for
those systems (For example, see Fig. 3 of Ref. \cite{Hiyama96} in the
case of the $^6$Li).
Regarding the calculation of three-nucleon systems such as the $^3$H and $^3$He nuclei, we take the Jacobian coordinates shown in Fig. 1 of Ref. \cite{hiyama}.

The Schr\"{o}dinger equation is given by
\begin{eqnarray}
( H - E ) \, \Psi_{JM,TT_z}(^{3,6,9}{\rm Z})  = 0 ,
\label{eq:schr7}
\end{eqnarray}
where
\begin{eqnarray}
&& \!\!\!\!\!\!\!\!\!\! \!\!\!\!\!\!  H=T+ 
       \!\sum_{\rm a,b} V_{\rm ab}
     \! + \!V_{\rm Pauli} ,
\label{eq:hamil7}
\end{eqnarray}
with the kinetic energy operator $T$, and $V_{\rm ab}$ is the interaction between constituent particles a and b [including the CP-odd hamiltonian of Eq. (\ref{eq:CPVhamiltonian})].
The OCM projection operator $V_{\rm Pauli}$ is given below for the $^6$Li and $^9$Be.
The total wave functions for the $^6$Li and the three-nucleon systems are described in Eqs. (3.2) and (66) in Refs. \cite{Hiyama96} and \cite{hiyama}, respectively.
The total wave function of the $^9$Be nucleus is described as 
\begin{eqnarray}
\Psi
&& \!\!\!\!\!\!\!_{JM, TT_z}(^9{\rm Be})
\nonumber \\
&&
= \sum_{c=1}^{2} \:
\sum_{nl, NL}
\sum_{IK} \sum_{sS}
C^{(c)}_{nl,NL}\: {\cal S}_\alpha
\Big[ \big[ \Phi ( \alpha_1 )  \Phi ( \alpha_2 )  \nonumber  \\
      &&  \times    
            [ \phi^{(c)}_{nl}({\bf r}_c)
         \psi^{(c)}_{NL}({\bf R}_c)]_\Lambda \, \chi_\frac{1}{2}(N_1) \big] \Big]_J.
     \label{eq:he7lwf}
\end{eqnarray}
Here the operator $\cal{S_\alpha}$ stands for the symmetrization between the two $\alpha$ clusters.
The spin function of the nucleon is denoted by $\chi_\frac{1}{2}$.

Following the Gaussian Expansion Method (GEM) \cite{Kami88,3nucleon,hiyama,Hiyama2012ptep}, we take the functional forms of $\phi_{nlm}({\bf r})$, $\psi_{NLM}({\bf R})$ as
\begin{eqnarray}
\phi_{nlm}({\bf r})
&=&
r^l \, e^{-(r/r_n)^2}
Y_{lm}({\widehat {\bf r}})  \;  ,
\nonumber \\
\psi_{NLM}({\bf R})
&=&
R^L \, e^{-(R/R_N)^2}
Y_{LM}({\widehat {\bf R}})  \;  ,
\end{eqnarray}
where the Gaussian range parameters were chosen according to the geometric progression:
\begin{eqnarray}
      r_n
      &=&
      r_1 a^{n-1} \qquad \enspace
      (n=1 - n_{\rm max}) \; ,
\nonumber\\
      R_N
      &=&
      R_1 A^{N-1} \quad
     (N \! =1 - N_{\rm max}).
\end{eqnarray}
The angular momentum space $l, L, \Lambda \leq 2$ was found to be sufficient to obtain good convergence of the calculated results.

The Pauli principle in the $N-\alpha$ and $\alpha - \alpha$ systems is taken into account in the orthogonality condition model (OCM) \cite{ocm}.
The OCM projection operator $V_{\rm Pauli}$ is represented by
\begin{equation}
V_{\rm Pauli}=\lim_{\lambda \rightarrow \infty}\sum_{f} 
\lambda \, |\phi_f(\mbox{\boldmath r}_{\alpha x}) \rangle  \langle \phi_f ( \mbox{\boldmath r'}_{\alpha x} ) |,
\end{equation}
which rules out the amplitude of the
Pauli-forbidden $\alpha - \alpha$ and $\alpha - N$ relative states $\phi_f(\mbox{\boldmath r}_{\alpha x})$ from the three-body total wave function \cite{Kukulin}.
The forbidden states are $f=0s, 1s, 0d$ for $x=\alpha$ and $f=0s$ for $x=N$, respectively.
The Gaussian range parameter $b$ of the single-particle $0s$ orbit in the $\alpha$ cluster $(0s)^4$ is taken to be $b=1.358$ fm to reproduce the size of the $\alpha$ cluster. 
We employ the $\alpha - N$ and $\alpha - \alpha$ interactions so as to reproduce the scattering phase shift of the $\alpha - N$ and $\alpha - \alpha$ systems at low energy \cite{kanada, hasegawa}.
Also we use the Argonne $v18$ nuclear force for the three-nucleon systems, and the Argonne $v8'$ interaction for the $N-N$ subsystem of the $^6$Li nucleus \cite{av18}.

\begin{figure}[htb]
\begin{center}
\includegraphics[width=8cm]{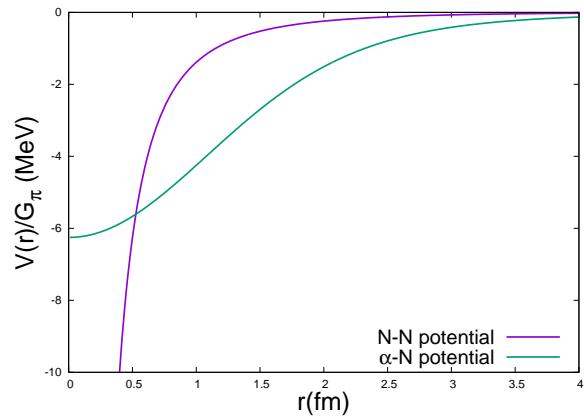}
\caption{\label{fig:folding}
Folding of the CP-odd one pion exchange $\alpha - N$ potential.
The CP-odd $N - N$ interaction is also shown for comparison.
The coupling constant $\bar G^{(1)}_\pi$ was factored out.
}
\end{center}
\end{figure}

For the CP-odd nuclear force, we have used the folding
\begin{equation}
V_{\alpha-N} ({\mathbf r})
=
\int d^3{\mathbf r'} \, V_{P\hspace{-.5em}/\, T\hspace{-.5em}/\, }  ({\mathbf r}-{\mathbf r'}) \rho_\alpha ({\mathbf r'})
,
\end{equation}
where $V_{P\hspace{-.5em}/\, T\hspace{-.5em}/\, }$ is the radial function of the CP-odd nuclear force and $\rho_\alpha (r) = \frac{4}{b^3 \pi^{3/2}} e^{-r^2 /b^2}$ is the nucleon number density of the $\alpha$ cluster normalized to 4.
The nucleon number density was approximated by a single gaussian with the spread $b=$ 1.358 fm.
The folding of the CP-odd potential is shown in Fig. \ref{fig:folding}.
It is important to note that the folding cancels the CP-odd nuclear forces for which the spin or isospin Pauli matrix acts on both interacting nucleons simultaneously.
Therefore, only the isovector pion exchange, isoscalar $\eta$ exchange, isoscalar and isovector $\omega$ exchange CP-odd nuclear forces survive for the $\alpha - N $ system.
For the $\alpha - \alpha$ system, all CP-odd nuclear forces cancel.

We must also note that the folding may overestimate the effect of heavy meson exchange $\eta, \rho, \omega$, since the $\alpha$ cluster approximation partially averages the microscopic level physics.
At the microscopic level, the nuclear force has a repulsive core which hinders the contribution of the short range CP-odd potential with heavy meson exchange.
This effect may break away when the folding averages the potential at the $\alpha$ cluster scale.
We will return again to this remark later in the analysis of the $^6$Li and $^9$Be EDMs.

\section{Results and discussion}

We now give the result of the evaluation of the EDM of the $^6$Li nucleus in the cluster approximation.
The result is shown in Table \ref{table:nuclearedm}.
For the isovector CP-odd one-pion exchange contribution, we have found that the $^6$Li EDM enhances the CP violation by 2 times compared with that of the deuteron.
The $^6$Li EDM is made of two components, namely the deuteron cluster polarization and the effect of the CP-odd $\alpha - N$ interaction.
The contribution from the deuteron subsystem to the $^6$Li EDM takes a value close to the single deuteron EDM (about 43\% of the total $^6$Li EDM).
This can also be seen by inspecting the $^6$Li EDM induced by the isovector exchange of $\rho$ and $\eta$ mesons, which has no contribution to the CP-odd $\alpha - N$ potential.
As the sensitivity to them takes values very close to those of the deuteron EDM, we may say that the deuteron cluster is relevant in the $^6$Li EDM.
The CP-odd $\alpha - N$ contribution to the $^6$Li EDM takes a comparable value as the contribution from the deuteron cluster.
This additional effect makes the enhancement of the $^6$Li EDM.

For the $\omega$ meson exchange CP-odd potential, the enhancement of the CP-odd effect in the $^6$Li is more accentuated.
We must however be careful on it, since it may be an overestimation, as the short range part was averaged due to the folding.
The calculated contributions of the isovector CP-odd $\rho$ and $\eta$ exchange potential are reliable since their effects arise only from the deuteron cluster.
We must also note that the isoscalar and isotensor CP-odd potentials do not contribute to the $^6$Li EDM, due to the inert $\alpha$ cluster and the isospin symmetry.
It is however to be noted that this is not the case if we consider the isospin breaking effect together with the excitation of the $\alpha$ cluster, since the spin and isospin structures become active \cite{chiral3nucleon}.
These topics are the subject of a more microscopic level study, and need to be investigated in future works.


We also give the result of the calculation of the $^9$Be EDM in the cluster approximation.
It is shown in Table \ref{table:nuclearedm}.
The sensitivity of the $^9$Be EDM to the isovector CP-odd pion exchange nuclear force is comparable to that of the deuteron.
Interestingly, the effect of the polarization due to the CP-odd $\alpha - N$ interaction is close to that for the $^6$Li EDM.

For heavier meson ($\eta , \rho , \omega$) exchange processes, the sensitivity of the $^9$Be on the CP-odd potential is much more important, but this is probably an overestimation, since all those effects arise from the dangerous folding of the heavy meson exchange CP-odd nuclear force.
As for the $^6$Li, there are also CP-odd interactions which may contribute with the excitation of the $\alpha$ cluster.
A more careful inspection of them are therefore needed at a more microscopic level.

\begin{table*}
\caption{
Result of the EDM calculation in this work.
The linear coefficients of the CP-odd $N-N$ coupling $a_X^{(i)}$ ($X=\pi , \rho , \eta , \omega $, $i=0,1,2$) are expressed in unit of $10^{-2} e$ fm.
The symbol $-$ denotes that the result vanishes in our setup.
The coefficient of the neutron EDM calculated in the chiral analysis \cite{crewther} was also added for comparison.
}
\begin{ruledtabular}
\begin{tabular}{l|cc|ccc|ccc|cc|cc|}
  &$\langle \sigma_p \rangle$ & $\langle \sigma_n \rangle$ &$a_\pi^{(0)}$ & $a_\pi^{(1)}$ & $a_\pi^{(2)}$& $a_\rho^{(0)}$ &$a_\rho^{(1)}$ &$a_\rho^{(2)}$ &$a_\eta^{(0)}$ &$a_\eta^{(1)}$ &$a_\omega^{(0)}$ &$a_\omega^{(1)}$ \\ 
\hline
\ $n$  & 0 & 1 & 1 & $-$ & $-1 $ & $-$ &$-$ &$-$ &$-$ &$-$ &$-$ &$-$    \\
$^{2}$H  & 0.914 & 0.914 & $-$ & $1.45 $ & $-$ & $-$ & $6.25 \times 10^{-2}$ &  $-$ &  $-$ & $0.157$ &$-$ & $-5.90 \times 10^{-2}$ \\
$^{3}$He  & -0.04 & 0.89 & $0.59$ & 1.08 & 1.68 & $-3.02 \times 10^{-2}$ & $4.26 \times 10^{-2}$ & $-7.68 \times 10^{-2}$ & $-5.77 \times 10^{-2}$ & 0.106 & $2.27 \times 10^{-2}$ & $-5.27 \times 10^{-2}$ \\
$^{3}$H & 0.88 & -0.05 & $-0.59$ & 1.08 & -1.70 & $3.07 \times 10^{-2}$ & $4.27 \times 10^{-2}$ & $7.86 \times 10^{-2}$ & $5.80 \times 10^{-2}$ & 0.106 & $-2.28 \times 10^{-2}$ & $-5.34 \times 10^{-2}$ \\
$^{6}$Li & 0.88 & 0.88 & $-$ & 2.8 & $-$ & $-$ & $7.1 \times 10^{-2}$ & $-$ & $-$ & 0.16 & $-$ & $-0.41$ \\
$^{9}$Be & $-$ & 0.45 & $-$ & $1.4$ & $-$ & $-$ & $-$ & $-$ & $-0.49$ & $- $ & $0.35$ & $-0.35$ \\
\end{tabular}
\end{ruledtabular}

\label{table:nuclearedm}
\end{table*}


We have also tested the ab initio calculation of the EDMs of the deuteron, $^3$He and $^3$H using the Argonne $v18$ interaction \cite{av18} and the CP-odd hamiltonian of Eq. (\ref{eq:CPVhamiltonian}).
The results are shown in Table \ref{table:nuclearedm}.
For the deuteron, the result is in good agreement with those of Refs. \cite{liu,song}.
We also point that the coefficients of the intrinsic nucleon EDM contribution $\langle \sigma_{p,n} \rangle_{^2{\rm H}}$ are smaller than one.
In the literature, these values were often set to one, but this is only true if the deuteron total angular momentum does not receive any orbital angular momentum contributions.
As the deuteron is known to have a $d$-wave component, the formula should be corrected.

Let us now present the result for the $^3$He and $^3$H nuclear EDM.
In our framework, the binding energies of the $^3$He and $^3$H nuclei obtained with the Argonne $v18$ nuclear force are $6.93$ MeV and $7.63$ MeV, respectively.
The difference from the experimental values (7.7 MeV for $^3$He and 8.4 MeV for $^3$H) is understood as the lack of the three-body force contribution \cite{3bodyforce}.
For the EDMs of the $^3$He and $^3$H, we have found values which are in agreement with the recent study using the chiral effective field theory \cite{bsaisou,bsaisou2}.
It is also in agreement with Ref. \cite{song} for the CP-odd isoscalar and isovector nuclear forces.
Our result differs from that of Ref. \cite{stetcu} by a factor of 1/2 for all CP-odd nuclear force contributions.


Let us finally see  the prospects for the observation of the new physics BSM.
If we model the new physics contribution by the exchange of new particles with mass $M_{NP}$ in the virtual state with $O(1)$ CP phase, the dimensional analysis gives the typical CP-odd $N-N$ coupling as $\bar G_\pi \sim g^2_{NP} \frac{\Lambda_{\rm QCD}^2}{M_{NP}^2} $, where $\Lambda_{\rm QCD} \sim 200$ MeV, and $g_{NP}$ is the coupling between quarks and new particles.
If the EDM of the $^6$Li nuclei can be measured at the level of $O(10^{-29})e$ cm, we thus can probe the new physics scale $M_{NP} \sim$ PeV [with $g_{NP} = O(0.1)$].
This na\"{i}ve estimation works for models which generate isovector CP-odd 4-quark interactions, such as the Left-right symmetric model \cite{Dekens:2014jka}.

For the supersymmetric model, the sensitivity of $O(10^{-29})e$ cm for the $^6$Li EDM can probe the CP phases of the $\mu $ term $\theta_\mu$ and the trilinear supersymmetry breaking coupling $\theta_A$ at the level of $O(10^{-2})$ for the supersymmetry breaking scale $M_{\rm SUSY} \sim$ TeV \cite{pospelovreview}.
Here we have assumed the Peccei-Quinn symmetry \cite{peccei}, and $\tan \beta = O(1)$.
The sensitivity to $\theta_\mu$ may be increased with growing $\tan \beta$ \cite{pospelovreview,demir}.
This prospective sensitivity can thus unveil the CP violation in the high scale supersymmetry breaking scenario.

We also give here the sensitivity to the class of models which generate Barr-Zee type diagrams.
This is the case for the Higgs doublet models \cite{2higgs}, supersymmetric models with R-parity violation \cite{yamanaka,rpv}, etc, and they contribute through the chromo-EDM \cite{cedm}.
Using the simple formula of the quark chromo-EDM $d^c_q \sim \frac{m_Q Y_q Y_Q}{16 \pi^2 m_{NP}^2} \ln \frac{m_Q^2}{m_{NP}^2}$, where $m_Q$ is the inner loop quark mass, $Y_q$ and $Y_Q$ are the couplings between the exchanged scalar and the quarks $q$ and $Q$, respectively, we obtain the sensitivity on the scale of new physics $M_{NP} \sim \sqrt{Y_q Y_Q} $ PeV.
If the coupling constants $Y_q$ and $Y_Q$ are small, the sensitivity to the scale BSM is attenuated.


\section{Summary}

In this paper we have calculated the EDM of $^2$H, $^3$He, $^3$H, $^6$Li, and $^9$Be nuclei using the Gaussian expansion method.
We have found that the $^6$Li enhances the CP-odd effect, due to the effect of CP-odd $\alpha - N$ interaction, in addition to the polarization contribution from the deuteron subsystem.
With the experimental sensitivity of $O(10^{-29})e$ cm for the $^6$Li EDM, we have strong chances to unveil the CP violation BSM at the PeV scale.
We therefore strongly recommend the experimentalists to study and measure the EDM of the $^6$Li nucleus.

\begin{acknowledgments}
The authors thank Prof. Koichiro Asahi, Dr. Jordy de Vries, Dr. Yasuro Funaki, Dr. Kaori Horii, Dr. Masahiro Isaka, Dr. Shota Ohnishi, Dr. Koichi Sato, Prof. Yannis K. Semertzidis, Dr. Hajime Togashi, and Dr. Yasuhiro Yamaguchi for useful discussions and comments.
NY also thanks the Workshop ``Structure and reaction of light exotic nuclei'' organized by the Yukawa Institute for Theoretical Physics.
The calculation was performed using the supercomputer HITACHI SR16000 of the Yukawa Institute for Theoretical Physics.
This work is supported by the RIKEN iTHES Project.
It is also supported by the HPCI Project.
\end{acknowledgments}

\end{document}